\documentclass[osajnl,twocolumn,showpacs,superscriptaddress,11pt]{revtex4-1} 
\usepackage{amsmath,amssymb,graphicx}
\pdfoutput=1 
\begin{document}

\title{Sub-nanometer flattening of a 45-cm long, 45-actuator x-ray deformable mirror}


\author{Lisa A. Poyneer}\email{Corresponding author: poyneer1@llnl.gov}
\affiliation{Lawrence Livermore National Laboratory, 
7000 East Avenue, Livermore, California 94550, USA}

\author{Thomas McCarville}
\affiliation{Lawrence Livermore National Laboratory, 
7000 East Avenue, Livermore, California 94550, USA}

\author{Tommaso Pardini}
\affiliation{Lawrence Livermore National Laboratory, 
7000 East Avenue, Livermore, California 94550, USA}

\author{David Palmer}
\affiliation{Lawrence Livermore National Laboratory, 
7000 East Avenue, Livermore, California 94550, USA}

\author{Audrey Brooks}
\affiliation{Northrop Grumman, AOA Xinetics Inc.,
115 Jackson Rd, Devens, MA 01434, USA}

\author{Michael J. Pivovaroff}
\affiliation{Lawrence Livermore National Laboratory, 
7000 East Avenue, Livermore, California 94550, USA}

\author{Bruce Macintosh}
\affiliation{Lawrence Livermore National Laboratory, 
7000 East Avenue, Livermore, California 94550, USA}

\begin{abstract}
We have built a 45-cm long x-ray deformable mirror
of super-polished single-crystal silicon that has 45 actuators along the tangential axis. 
After assembly the surface height error was 19 nm rms.
With use of high-precision visible-light metrology and precise
control algorithms, we have actuated
the x-ray deformable mirror and flattened its entire surface to 
 0.7 nm rms controllable figure error.
This is, to our knowledge, the first sub-nanometer active flattening
of a substrate longer than 15 cm.
\end{abstract}


\maketitle 


\section{Introduction}
\label{sec:intro}

The advent of 4th-generation x-ray light sources (i.e., free electron 
lasers like the Linac Coherent Light Source in the U.S. and 
SPring-8 Angstrom Compact free electron laser in Japan 
and advanced synchrotrons like the National Synchrotron Light 
Source II in the U.S.) requires increasingly advanced and 
high-performance x-ray mirrors.  Combining expertise in visible 
wavelength adaptive optics and reflective x-ray optics, Lawrence 
Livermore National Laboratory (LLNL) has begun a research and 
development effort to design, fabricate and test x-ray deformable mirrors.

X-ray deformable mirrors could provide two significant
benefits over traditional non-adaptive x-ray optics. First, active control 
is a potentially inexpensive way to achieve better surface figure than
is possible by polishing alone, particularly on long substrates. 
Secondly, the ability to change 
the figure allows for dynamic correction of aberrations in a 
x-ray beam line. This includes both self-correction of errors in the mirror itself (such as
those caused by thermal loading) and correction of errors on other optics, 
the latter of which has been demonstrated elsewhere\cite{Mimura2010Breaking-the-10}.

With these goals in mind we have built a 45-cm x-ray deformable mirror
(XDM).  As detailed below, this mirror was designed to provide
fine-scale control of its surface. Using precise visible-wavelength
metrology, we have been able to generate voltage commands for the
XDM's actuators that flatten it to as good as 0.7 nm rms, which is
significantly better than the initial substrate polishing before
assembly. The following sections describe the XDM, the metrology
equipment, our calibration and control methods and finally the
flattening results.

To place our XDM in context, we must consider two types of mirrors
that have been developed by others in the field. The first is with
non-active super-polished mirrors. We need to be able to control our
XDM to a comparable flatness. Our XDM was designed with the same size
specifications as the hard x-ray offset mirrors (known as HOMS) for
LCLS~\cite{doi:10.1117/12.795912,Barty2009Predicting-the-}. Visible-light
metrology (using the same interferometer that we have used for this
work) on the four delivered HOMS measured the figure errors (which
exclude cylinder) between 1.0 and 2.4 nm
rms~\cite{doi:10.1117/12.795912,Barty2009Predicting-the-}. JTEC
produces mirrors up to 50 cm long, with a claimed shape error of less
than 0.5 nm rms at best effort~\cite{JTECwebsite}.

Although fixed figure 0.5 m flats with $<$ 0.5 nm rms error are being
produced, the advantages of a variable figure capable 0.5 nm figure
error tolerance motivate this study. For example, a variable figure
mirror can compensate for localized heating that drives figure error
well above the as manufactured specification. If mirrors are coated, a
concomitant cylinder can be compensated. Finally, a deformable mirror
may aid in compensating figure errors introduced by final focusing
optics, which are not yet being manufactured to diffraction limited
performance. One deformable mirror can correct the net sum of all
these effects, without the need to fully understanding the origin of
each component.

The second point of comparison for our XDM is to other deformable x-ray optics.
Below, we summarize published performance of the best flattening
achieved for other deformable x-ray mirrors.
In 2010 a French collaboration~\cite{Mercere2010Hartmann-wavefr} 
developed an active x-ray mirror to be deployed at the SOLEIL facility; 
this mirror implements a 35 $\times$ 4 $\times$ 0.8 cm silicon 
substrate held between an active jaw and a flexor, to generate variable 
elliptical profiles. In addition the mirror features 10 actuators across its 
length to minimize asphere. The actuators are perpendicular to the mirror 
surface, and force is applied by a spring-floating head coupled to a stepper 
motor. Actuator hysteresis was reported to be 0.1$\%$.  The SOLEIL team 
demonstrated flattening of the mirror down to 3.0 nm rms of asphere, 
and 0.6$\mu$rad slope errors over 30 cm of clear aperture, with a 
maximum radius of curvature equal to 60 m.

In the same year, the Diamond Light Source developed a 15 $\times$ 
4.5 cm adaptive x-ray mirror, including eight piezo bimorph 
actuators~\cite{Sawhney2010A-novel-adaptiv}. This mirror was specifically 
designed to achieve a high level of figure control, while allowing for adjustable 
radius of curvature. The actuated mirror, built by SESO and super polished by 
JTECH via Elastic Emission Machining (EEM), reached 0.66 nm rms of asphere 
over 12 cm of clear aperture, with the smallest beam size at 
focus equal to 1.2 $\mu$m FWHM.

Another x-ray deformable mirror was developed in Japan and deployed at
SPring-8 before a pair of Kirkpatrick-Baez (KB) mirrors, to achieve
nearly diffraction-limited beam
focusing~\cite{Mimura2010An-adaptive-opt}. The 12 cm silicon substrate
was super-polished by EEM, and features 16 piezoelectric
plates. During operation, a Fizeau interferometer was placed in front
of the mirror to provide real time figure correction. The experimental
team reported a 7 nm FWHM spot size at the focus plane. For this
experiment an estimated 10 nm peak-to-valley surface profile was found
in situ, but several error sources were listed. Previously, visible
light metrology on this optic~\cite{Kimura2008Development-of-} outside
of the beam line was conducted with Fizeau interferometry. An
approximately 2 nm peak-to-valley figure error was measured at best
flat. No rms figure error was reported; for a typical figure error PSD
this is approximately 0.7 nm rms.
 
In 2012 the x-ray optic group at the Elettra Synchrotron facility in Italy 
reported on the successful construction of an adaptive x-ray mirror for the 
TIMEX beamline at FERMI~\cite{doi:10.1117/12.929701}. This mirror is 
40 cm long and 4 cm wide, and features 13 piezo actuators and 13 
strain gauges. Rough flattening of the mirror is first achieved by acting on 
four clamps located on the mirror mount. The idea of including calibrated 
strain gauges in the design allows the mirror to work in closed loop without 
the need of a wavefront sensor. The same idea was implemented for the 
design of our XDM. To our knowledge no precision flattening
results have been reported from Elettra.

\section{Deformable mirror design}

Our single crystal silicon substrate is 45 cm in length, 3 cm high,
and 4 cm in width. Substrate quality is discussed in
Section~\ref{sec:substrate}. The mirror and actuators are supported on
an invar mount, and enclosed in a protective housing that leaves the
reflective surface exposed to grazing incidence x-rays.  
The
substrate was cut from a large boule similar to those typically used
in wafer fabrication, which are usually pulled in the (111)
direction. Typical resistivity is $<$ 10 ohm-cm. Neither parameter has
any significant affect on our mirror's performance. As of now we do
not expect to deposit any single or multilayer coating on the silicon
substrate. Preliminary experiments at a synchrotron facility will be
conducted at low photon energy and grazing incidence, well within the
critical angle of silicon.

\begin{figure}[htbp]
\centerline{
\includegraphics[width=\columnwidth]{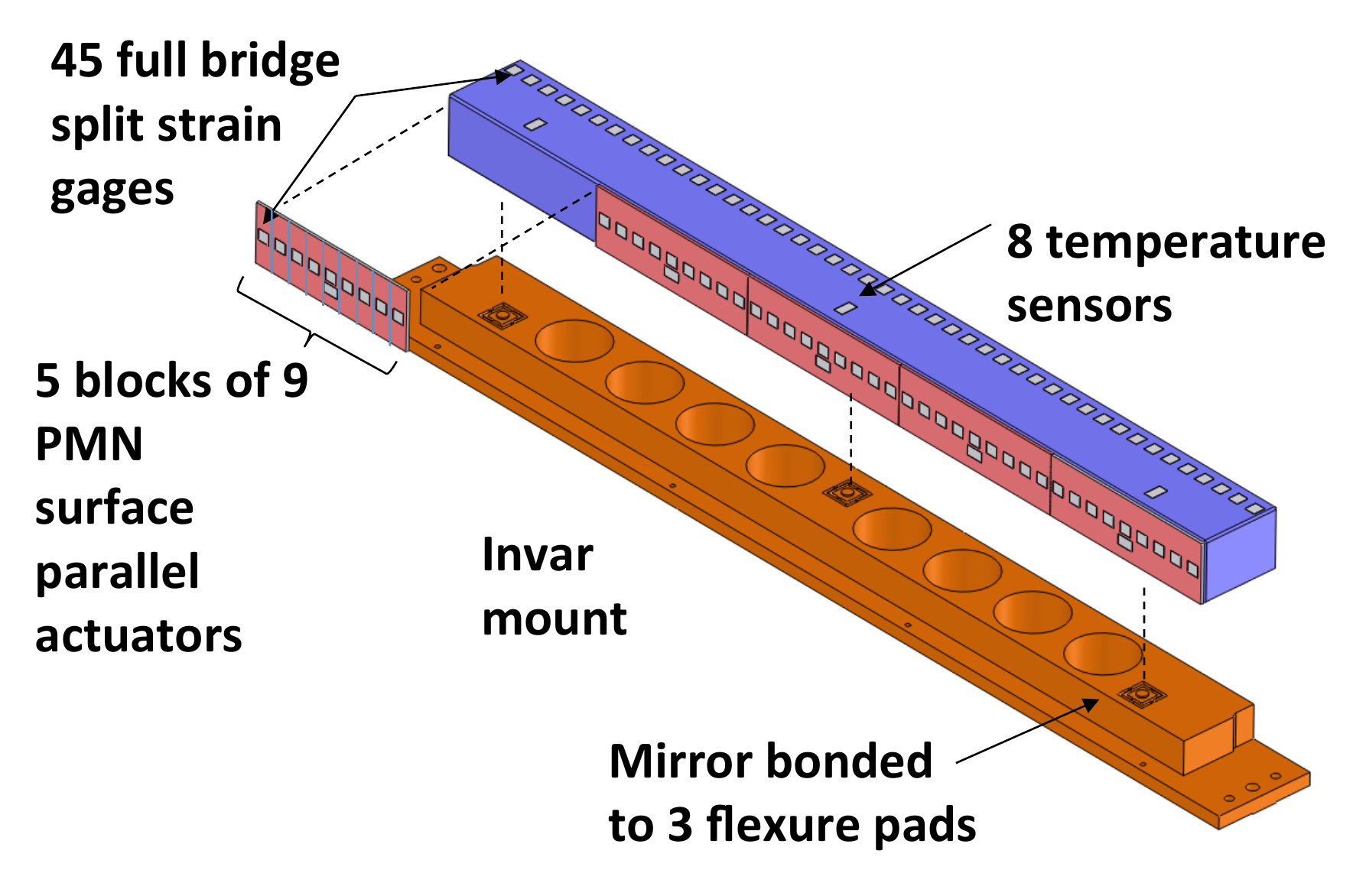}
}
\centerline{
\includegraphics[width=\columnwidth]{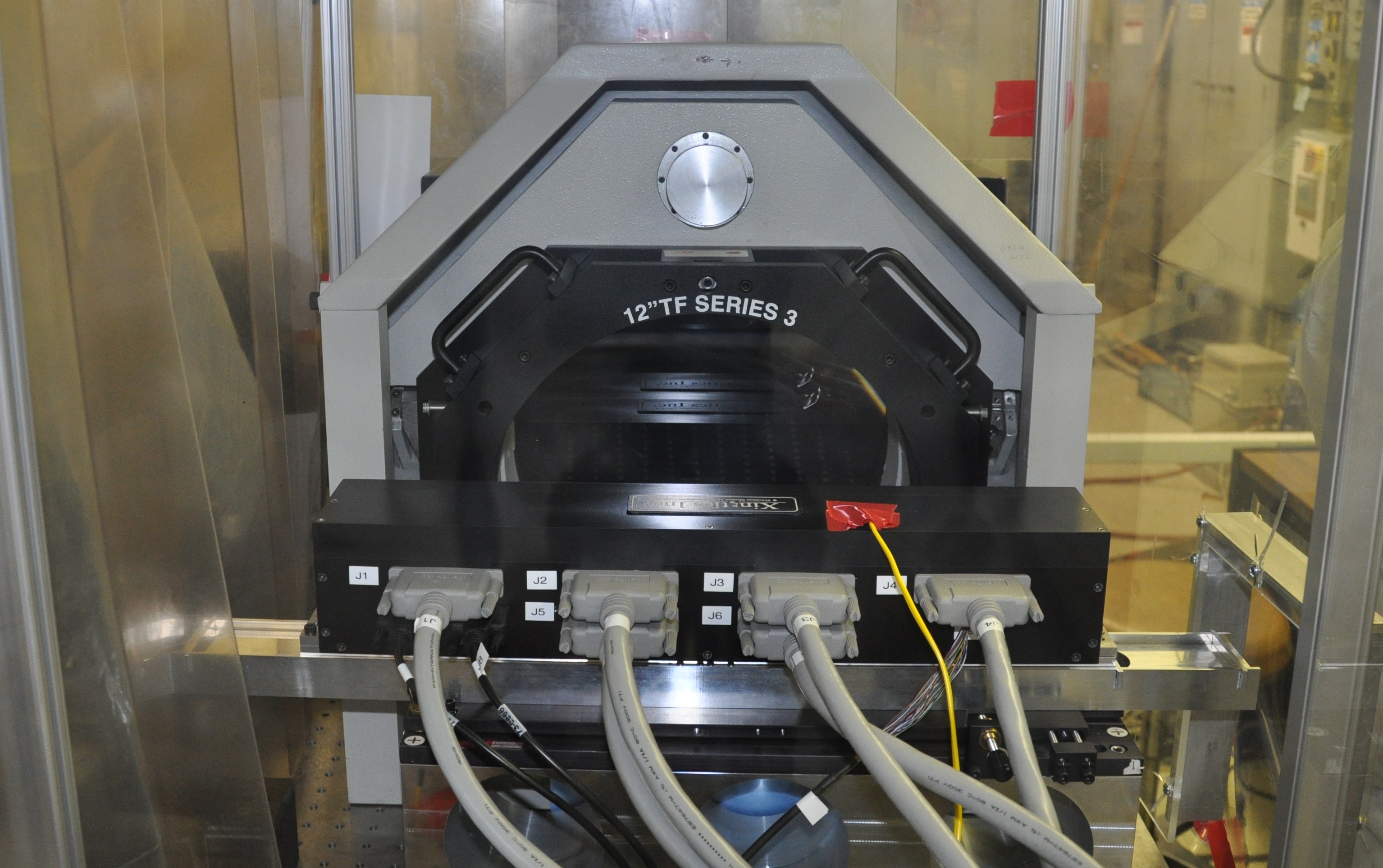}
}
\caption{[Top]: The XDM has 45 parallel actuators along the back
  surface, 45 full-bridge strain gauges at the top, and 8 temperature
  sensors. It is mounted with three flexure pads to isolate the XDM's
  motion. [Bottom]: photograph of the XDM facing the interferometer in
  our metrology laboratory, courtesy of Jeff Bonivert.}
\label{fig:fig_solid}
\end{figure}

Figure~\ref{fig:fig_solid} illustrates how the 45 actuators are bonded on the side 
opposite the reflective surface. 
Each actuator is 1 cm long, 3 cm 
high, and about 0.15 cm thick. They are spaced evenly every 1 cm along the
tangential axis of the mirror. The surface parallel actuator geometry of the actuators, 
along with three flexure supports machined into the Invar mount, 
minimize unintended forces on the mirror, and their  effect on figure during 
actuation. The actuators are epoxy bonded to the mirror while at 
mid-range of their operating voltage, which enables the mirror to bend both 
concave and convex. 
The mirror is bonded to the three flexure pads to constrain all motion 
except that induced by the actuators. The flexures also 
isolate the mirror from differential thermal expansion between the substrate and mount.

The actuators operate from 15 to 75 volts. The actuators were bonded to the 
substrate at 45 V. This enables the XDM to make spherical surface in either direction
from a nominal flat surface. The voltage at which the mirror has no curvature, nominally 45 V, is referred
to as the bias voltage.

Figure~\ref{fig:fig_solid} also illustrates the location of 45 full bridge strain gauges. 
One half of each bridge is bonded to the back side of each actuator. 
The mating half-bridge is bonded to the top of the mirror, where 
strain is similar to that at the mirror's reflecting surface. The strain resolution 
of each gauge is 
about 10 parts per billion (commonly referred to as nanostrain). This corresponds to 
each gauge measuring surface figure changes to better than 
1 nm between gauges.

 The strain gauges can detect and  correct for 
differential expansion between the mirror and actuators. However, 
because the gauge response itself may be slightly temperature 
sensitive, Resistance Temperature Detectors (RTDs) are bonded to 
the mirror to measure temperature: 
three on the top of the mirror, and one to each of the five 
Lead Magnesium Niobate (PbMnNb or PMN) 
actuator blocks. The eight RTD locations are also shown in Figure~\ref{fig:fig_solid}.

The mirror's actuators, strain gauges, 
and RTD's are all wired to a printed 
circuit board  that is secured to the back of the 
mirror mount. Each wire from the mirror is soldered to the printed circuit 
board. Electrical connectors embedded in the board  connect the mirror assembly to 
power supplies and signal processors. The circuit board, mirror, and mount after
assembly are enclosed, with cables required for operation connected to the back. 

Currently the mirror is compatible with operations in vacuum limited
to $10^{-7}$ mbar; this should offer enough flexibility in terms of mirror
deployment, especially considering that we can always take advantage
of differential pumping, if an ultra-high vacuum environment is
needed. Also, we do not expect appreciable temperature changes on the
mirror as a consequence of x-ray heat load, since the shallow grazing
incidence design of the optic guarantees a large footprint of the
x-ray spot.

In order to provide very high quality strain gauge readings, while
also keeping costs in line, the 45 strain gauge signals coming off the
XDM printed circuit board are multiplexed into a single channel of an
MGCplus measurement system (HBM, Inc).  The MGCplus is outfitted with
an ML38B amplifier and conditioning module.  The signals are
multiplexed using an Agilent 34980A data acquisition box with three
34922A multiplexer modules.  Although cost and simplicity are benefits
of this approach, a detriment is that we need to wait a considerable
amount of time between strain gauge readings to let a long time
constant low-pass filter settle each time the multiplexer is switched.
At present, we wait 20 seconds between reading each strain
gauge, for a total of 15 minutes for all 45 gauges.  The 8 RTDs are
multiplexed through the same Agilent 34980A and read with an Agilent
34411A DMM.  Much less conditioning is required for the RTDs and,
hence, they can be read almost instantaneously after a multiplexer
switch.  The 45 actuators are controlled using Northrup Grumman USB DM
drive electronics.  All of these electronics are connected to the
control computer over serial channels. For ease of development and
flexibility during testing, Matlab was selected as the software
development environment.  Software developed by the team provides the
control, measurement, and analysis capabilities needed to support the
work described herein.

Though we do not use the strain gauges and RTDs in the work described
here, we describe them for completeness. These sensors were included
to help us control the XDM's stability through time and with
temperature changes, which is a subject for future work.  We next
discuss the visible light metrology that we use to characterize the
XDM.

\section{Visible-light metrology}

\subsection{The 12-inch Zygo interferometer}

The Zygo Mark II phasing interferometer used in this work 
has a noise floor of about 0.3 nm, and is calibrated to 
measure figure with an absolute accuracy approaching 
1 nm (rms) over the 28 cm field of view. The 45 cm 
deformable mirror figure is constructed by stitching three 
28 cm long interferograms. Without suitable characterization, 
interferometer calibration errors will produce inconsistencies within stitched regions, 
limiting the ability to demonstrate deformable mirror performance.

\begin{figure}[htbp]
\centerline{
\includegraphics[width=\columnwidth]{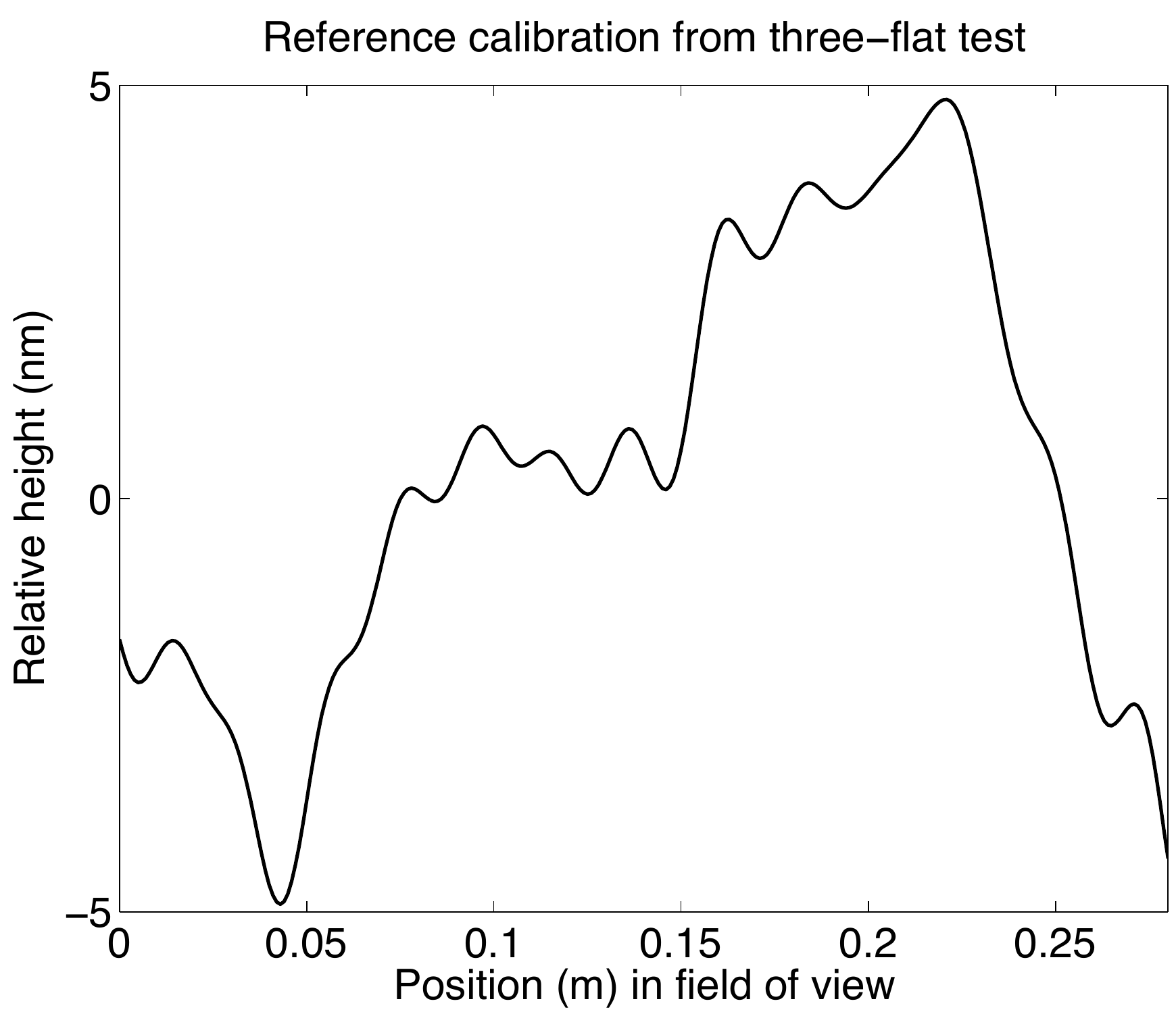}
}
\caption{The reference calibration for the interferometer,
as determined by  three-flat test. This shape is subtracted from raw measurements
to convert relative height to absolute height.}
\label{fig:fig_refflat}
\end{figure}

Therefore a three flat test was used to calibrate the interferometer. 
During the test two transmission flats, T1 and T2,
are mounted onto the interferometer, and also placed at the optic 
under test location.  The third optic is a reflection flat labeled R. 
It is placed at the optic under test location, and is 
rotatable 180 degrees about its optic axis. 
The figure of each optic along a horizontal line can be calculated from the three data files,
and the solution for T2 is used as the reference calibration when measuring 
our XDM. This reference calibration is shown in Figure~\ref{fig:fig_refflat}.
This amplitude of the correction is $\pm5$ nm, significantly more than the signal that
we want to measure at best flat.

This study required over a hundred surface figure measurements per day
during the course of algorithm optimization. This measurement rate
would best be met with a full aperture visible light
interferometer. However, the 600 mm aperture interferometers tested
had a measurement noise floor well above the 0.1 nm noise floor of our
best performing 300 mm aperture unit. Off-normal incidence angles with
additional reflectors were tested using the 300 mm unit, but the
additional air path length severely increased the intrinsic instrument
noise. As a result, the 300 mm unit at normal incidence was selected,
and stitching employed to measure the full aperture. The performance
of this method in measuring 450 mm mirrors compared favorably with
long trace profile measurements made at
LBL~\cite{doi:10.1117/12.945915}, as well as in-situ measurement made
at x-ray wavelengths at LCLS~\cite{Rutishauser2012Exploring-the-w}.

\subsection{Mounting and measuring the XDM}

As noted above, we take three measurements of the XDM surface and
stitch them together. The mirror's three positions in front of the interferometer are termed 
right, center and left. The right position
corresponds to the lowest numbered actuators on the XDM. The center
position is approximately centered on actuator 23. The left position corresponds to the 
higher numbered actuators on the XDM. This three-measurement setup
provides 20 cm overlap in the two stitching zones.

The centerline of the XDM is matched to the calibrated horizontal line 
of the interferometer. The mount is moved from right to 
center to left with no vertical motion to ensure the interferometer 
is measuring the mirror at the same elevation.
The mount is designed with stops to ensure repeatability of better than 1 mm
(which is one pixel in the interferometer, see below) as it is moved.

Interferometer measurements are mapped to the physical 
surface of the XDM by adjusting interferometer magnification to 
1 mm of mirror surface/pixel. Each measurement
is 288 pixels long, corresponding to 28.8 cm on the XDM surface. 
We define an x-axis along
the centerline of the XDM, with $x=0$ at actuator 23. In each 
of the three positions different actuators are bent and measurements 
are taken to determine the exact portion of the XDM that
the interferometer measures.
At present, in the right position the measurement spans  -22.2 to 6.5 cm
along the x axis (as defined above); in center position it spans -14.1 to 14.6 cm;
in left position it spans -6.6 to 22.1 cm.

\begin{figure}[htbp]
\centerline{
\includegraphics[width=\columnwidth]{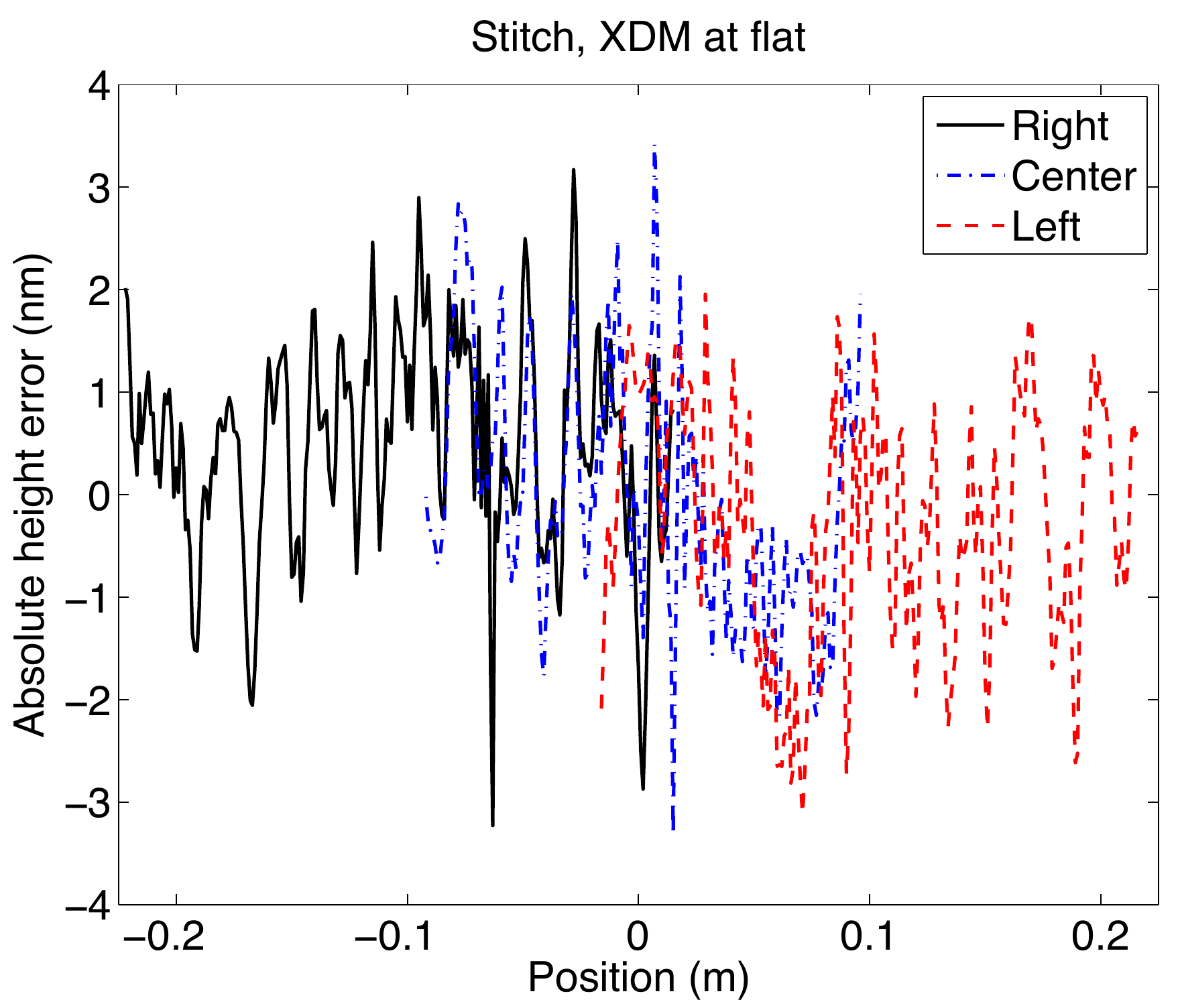}
}
\caption{Calibrated interferometer measurements
stitch very well. Three measurements were taken at the same voltage, for 
right, center and left positions. }
\label{fig:fig_stitch_flat}
\end{figure}

Calibration accuracy is essential for stitching interferograms. 
The interferometer calibration file is subtracted from 
each measurement to yield the absolute height
of the XDM.
 Then piston (constant height) and tilt (linear height) are removed from each measurement.
Then, ignoring the fifty pixels at either end of the measurements, we align the
remaining overlap between the right and center measurements. 
This alignment is done
by adjusting the tilt and piston on the right measurement to produce
the minimum squared error between it and the center measurement in 
the overlap region. Then we repeat the procedure to align the left measurement
to the center, again minimizing the squared error. The results of such an 
alignment are shown in Figure~\ref{fig:fig_stitch_flat}. 
The excellent
agreement of the three measurements verifies the quality of the calibration
via the three-flat test. The final stitched measurement is produced by 
taking the mean of all valid sample points (from either one, two or three 
positions) for each x location. Points at the ends of the lineouts are ignored
if they display artifacts of miscalibration. The resulting stitched measurement
is always absolute height.

A stitched measurement represents nearly the entire 45-cm length of
the XDM. Figure~\ref{fig:fig_stitch_flat} shows very good agreement
between the different views of the same mirror shape, for there
measurements taken with the mirror held at a fixed voltage (from one
of the flattening experiments, see Section~\ref{sec:fullflat}).  Since the
calibration by the three-flat test is $\pm 5$ nm (see
Figure~\ref{fig:fig_refflat}) this excellent agreement gives us
confidence that the calibration is correct.  All claims about figure
error are made relative to this calibration. Of course, the
calibration may be slightly wrong, and hence the flattening not quite
as flat. However all external metrology, whether of deformable or
static x-ray optics, will require calibration. If when testing at a
x-ray light source our best flat produces an x-ray beam of lower than
expected quality, we will be able to change the voltages commanding it
to improve the figure, dependent accurate calibration of any in situ
metrology.

Also apparent in Figure~\ref{fig:fig_stitch_flat} is that there is
significant high-spatial frequency content in the interferometer measurements. Though some
of this may represent high-frequency polishing errors on the XDM, most of
it is noise. In the literature~\cite{doi:10.1117/12.795912,Sawhney2010A-novel-adaptiv}
such noisy measurements are usually low-passor median filtered.

The fundamental limits of phasing interferometer noise are discussed
in~\cite{Sommargren2002100-picometer-i}. The noise floor of our
measurements correspond to about 632 nm/1000, which is found by many
researchers~\cite{Malacara1992Optical-Shop-Te} to be the performance
limit for commercially available equipment. This performance is only
achieved when environment vibration and air turbulence are fully
suppressed, leaving only the intrinsic noise of the measurement
machine. If there were a single cause to this floor, it could be
addressed and corrected by interferometer manufacturers.

In our case we have a natural characteristic frequency for the system that
is set by the XDM. Since the actuators are spaced every one centimeter, 
the highest controllable mode has a period of two centimeters. The XDM
cannot make shapes of higher spatial frequency. 
When assessing the performance of our flattening, we only consider the spatial
frequencies below this cutoff. To obtain this portion of the signal, we simply low-pass filter the
measurement with a hard cutoff in our software. We term such a filtered measurement 
as the controllable height. All results presented below will 
quote flattening performance in terms of the controllable height.

There is also a meaningful distinction to be made between the complete
measurement of the XDM's surface and its spherical and aspheric components.
This distinction is typically made (see Section~\ref{sec:intro}) in the literature. 
In our case the component due to curvature of the surface is termed cylinder,
and represents a height that is a quadratic function of the x-position on the mirror.
For our mirror this cylinder is controlled by changing the average values of the 
actuators voltages. As noted above, there is nominally no cylinder at
the bias voltage of 45 V. However, the amount of cylinder
varies with temperature.
Further characterization and control of this is left for future work. 
For the purposes of this work, we minimize the cylinder but disregard any
small change that may have crept in during the execution of our experiments.

\section{Substrate characterization}
\label{sec:substrate}
The single-crystal silicon substrate was produced by InSync, Inc 
(Albuquerque, NM) and polished by QED Technology (Rochester, NY) 
via Magneto-Rheological finishing (MRF). Upon receipt, 
extensive characterization of the mirror surface was conducted at 
LLNL, including atomic force microscopy (high-spatial frequency 
roughness), white light interferometry (mid-spatial frequency roughness), 
and large aperture interferometry (figure error). 

The surface roughness at the center of the mirror in the high-spatial frequency 
range of 0.33$\mu$m$^{-1}$ -  50$\mu$m$^{-1}$
(often referred to as ``finish'') was measured to be 3.7 \AA, close to the specification 
of 4.0 \AA. The roughness at the center of the mirror in the mid-spatial frequencies 
of 10$^{-3}$ $\mu$m$^{-1}$  -  33$\mu$m$^{-1}$ (often referred to as ``mids'')
was 5.2 \AA; this is above the specification
of 2.5 \AA. The roughness in the mids was dominated by the lowest
spatial frequencies.
The power spectral density was computed by stitching data from these 
measurements to cover both the mids and finish. It follows the expected fractal 
behavior described by Church et al.\cite{Church1988Fractal-surface}
\begin{figure}[htbp]
\centerline{
\includegraphics[width=\columnwidth]{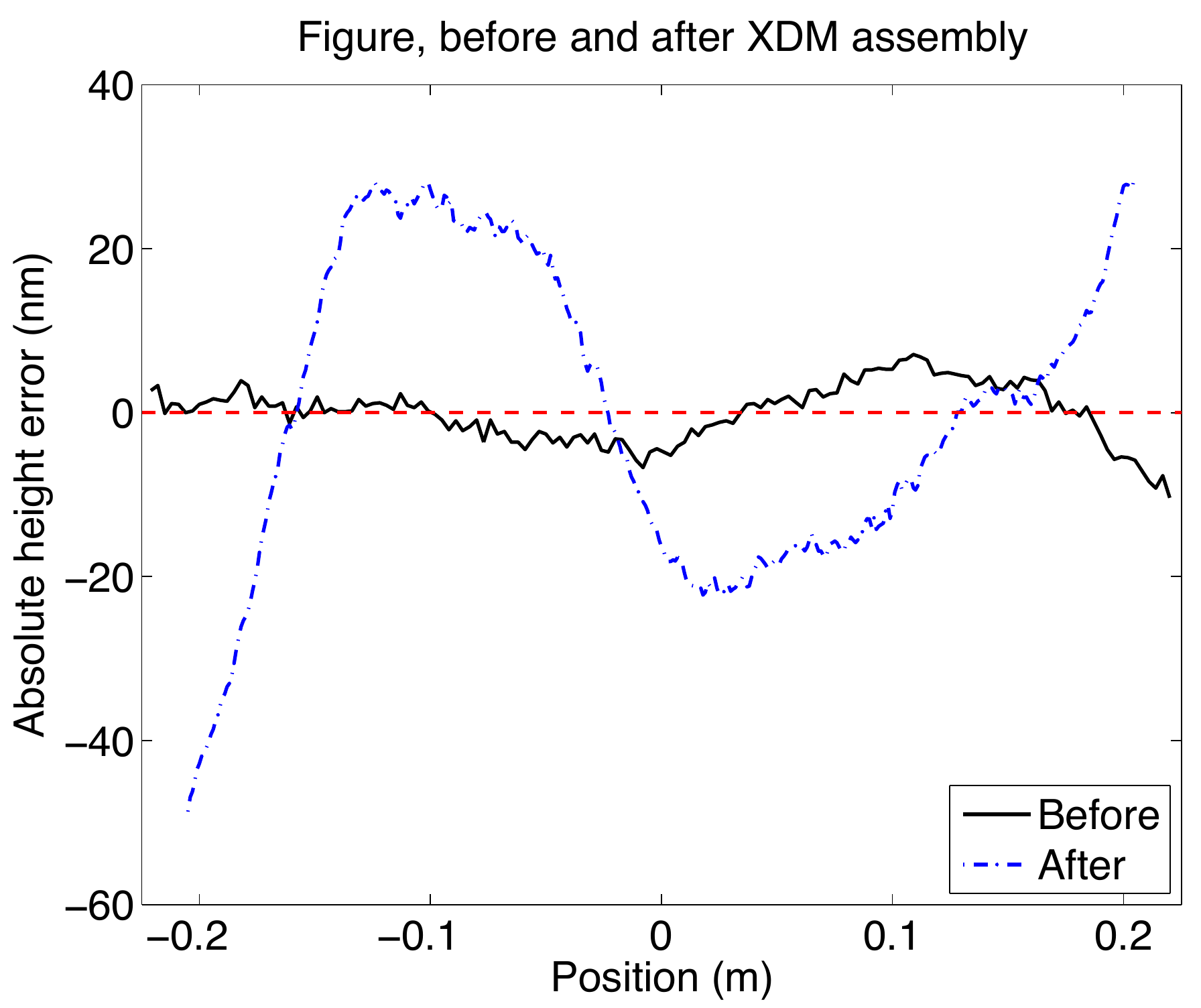}
}
\caption{After polishing the substrate's figure 
error was 3.5 nm rms. After actuator bonding and mirror assembly, the figure
error had increased to 19 nm rms, with significant low-frequency figure errors.
Stitched interferometer measurements
are shown.}
\label{fig:fig_bef_aft_sub}
\end{figure}
The substrate's figure was measured with the interferometer described above,
both before and after actuator bonding and mirror assembly. 
Figure~\ref{fig:fig_bef_aft_sub} presents both measurements. 

After polishing and
before assembly the figure error was 3.5 nm rms. After assembly
at uniform voltage the figure error was 19 nm rms. This amount is well within the
dynamic range of the mirror to self-correct (as it was designed to be). 
To remove this 19 nm rms figure error, we must determine the proper voltages.

\section{Characterization and control of the deformable mirror}

Just as in astronomical or vision-science adaptive optics, the challenge of
controlling the XDM is to determine the set of commands that 
produce a desired  shape on its surface.  In our experimental
setup we have a very high-quality height measurement of the surface.
Given this 
height, we must ``fit'' it to the deformable mirror by determining 
the  set of actuator commands that best corrects that shape. 
(See Ellerbreok~\cite{MVU} for a 
thorough discussion of this concept in the field of astronomical adaptive optics).

If our XDM is a linear system (which it approximately is), we can
describe it with a simple matrix equation.
Given a vector of 45 voltages $\mathbf{v}$, the height $\mathbf{\phi}$ made on the 
surface of the XDM follows the matrix equation
\begin{equation}
\label{eqn:forward}
\mathbf{\phi} = \mathbf{H}\mathbf{v},
\end{equation}
where the matrix $\mathbf{H}$ describes the response of the XDM.
In this case the height $\mathbf{\phi}$ has the same sampling
and number of pixels as the stitched interferometer measurements.
Then, given a desired height shape on the XDM, we can ``fit'' the 
height and estimate the voltages by solving the inverse problem.
In the following subsections we discuss how to obtain $\mathbf{H}$,
if the underlying assumptions of the linear model are true, and 
how best to go about solving the inverse problem given the 
unique characteristics of the XDM.

As noted above the mirror is commanded around a non-zero bias voltage
which produces a surface with no cylinder. So for clarity in
notation, for the remaining treatment assume that the vector
$\mathbf{v}$ represents the voltage value relative to bias, as opposed
to the actual voltage commanded through the electronics.

\subsection{Influence function}

The term influence function refers to the shape that the XDM makes
in response to voltage applied to a single actuator. During the development and
design of the XDM, a detailed finite-element-analysis model was constructed. It
produced the estimated influence function for each of the 45 actuators on the 
XDM, sampled at 1 mm per pixel. By taking the output of the FEA model
along the centerline of the XDM, we can populate the matrix $\mathbf{H}$,
with each column representing the height made by one actuator.
The XDM cannot make tilt across its full length, and the influence functions 
contain no tilt. They are furthermore offset to have no piston (average value),
which cannot be measured by the interferometer and is irrelevant to the
wavefront error.

\begin{figure}[htbp]
\centerline{
\includegraphics[width=\columnwidth]{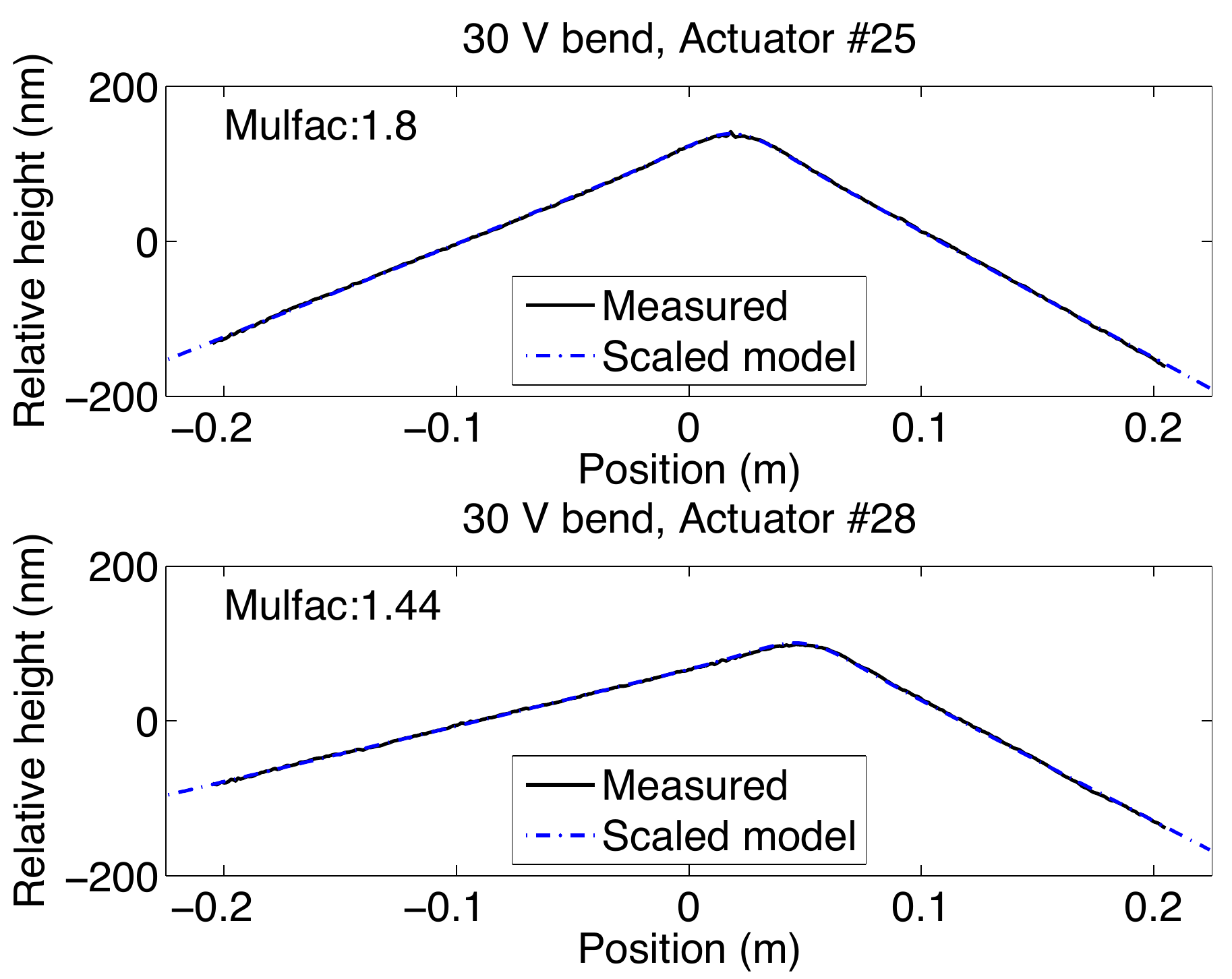}
}
\caption{The actual response of the XDM to single-actuator motion agrees very well
with finite-element modeling, with a scaling factor adjustment for total motion.
Most actuators, like actuator 25 (top), produce 1.8 times more motion than
the model predicted. Four actuators, including actuator 28 (bottom), 
bent less than the rest.}
\label{fig:fig_acts}
\end{figure}

Just such a matrix was used for our initial control of the XDM. To
determine how accurate the model was, we performed an actuation
test. The mirror was commanded to bias voltage, and then one actuator
was commanded to 30 V above bias, or half the total voltage
range. This pair of moves was repeated for all 45 actuators. This
entire process was done in each of the mirror's three mount
positions. To analyze the data, the measurement at bias was subtracted
from the measurement when an actuator was commanded to produce a
change in surface figure. (This is necessary to remove the figure of
the XDM at bias voltage, which the FEA model does not know about.)
Finally, for each actuator the measurements at the three mount
positions were stitched together. The response of actuator 25 is shown
at top in Figure~\ref{fig:fig_acts}.
As with all actuators, the shape agrees very well with the model but
the magnitude is higher than the FEA predictions. The parallel
actuator configuration utilizes the in-plane strain of the PMN for
actuation therefore the stroke of the PMN actuators can not be
directly measured before assembly. To guarantee that the stroke
requirements are met, the XDM was designed with a conservative
actuator strain learned from heritage AOX mirrors. A scaling factor of
1.8 was used to match the magnitude of the conservative FEA result
with the actual measured stroke.

This analysis was done for all 45 actuators. Each actuator has a unique
influence function - actuators closer to the edge have less displacement.
Most actuators had this same 1.8 scaling factor but four did not.
Actuator 28 is shown at bottom in Figure~\ref{fig:fig_acts}. For this actuator
the shape agreement is excellent, but now the scaling factor is 1.44.
This means actuator 28 does not respond as much to the same voltage as
most other actuators.

After this full characterization, we modified the initial $\mathbf{H}$ that 
was based on the FEA model. Most columns were multiplied by 1.8 to 
reflect the actual behavior of the XDM. Actuators 27, 28, 36 and 37 were 
given different scaling factors based on the analysis described as above. This 
produced our final $\mathbf{H}$ matrix.

Given this model, we can evaluate the dynamic range of the XDM as
built. By commanding all actuators uniformly to either maximum or
minimum voltage, we can make 7.4 microns peak-to-valley cylinder in
either direction. Due to the broadness of the influence function,
stroke goes down rapidly with spatial frequency. The model indicates
that the XDM can make 750 nm peak-to-valley of a sine wave of two
cycles across the 45 cm length, but only 40 nm peak-to-valley of a
sine wave of eight cycles. This is more than sufficient to
self-correct the mirror, as we demonstrate below.

\subsection{Verification of linearity}

The fundamental assumption behind Equation~\ref{eqn:forward}
is that the XDM behaves as a linear system. This requires two things\cite{OpWil}.
First, given two different inputs $\mathbf{v}_1$ and $\mathbf{v}_2$
that produce outputs $\mathbf{\phi}_1$ and $\mathbf{\phi}_2$,
 the sum of the inputs  $\mathbf{v}_1 + \mathbf{v}_2$ must
produce an output that is equal to the sum of the individual outputs
$\mathbf{\phi}_1 + \mathbf{\phi}_2$. This property of linear superposition
holds true for piezo-actuated DMs made previously by Xinetics.

\begin{figure}[t]
\centerline{
\includegraphics[width=\columnwidth]{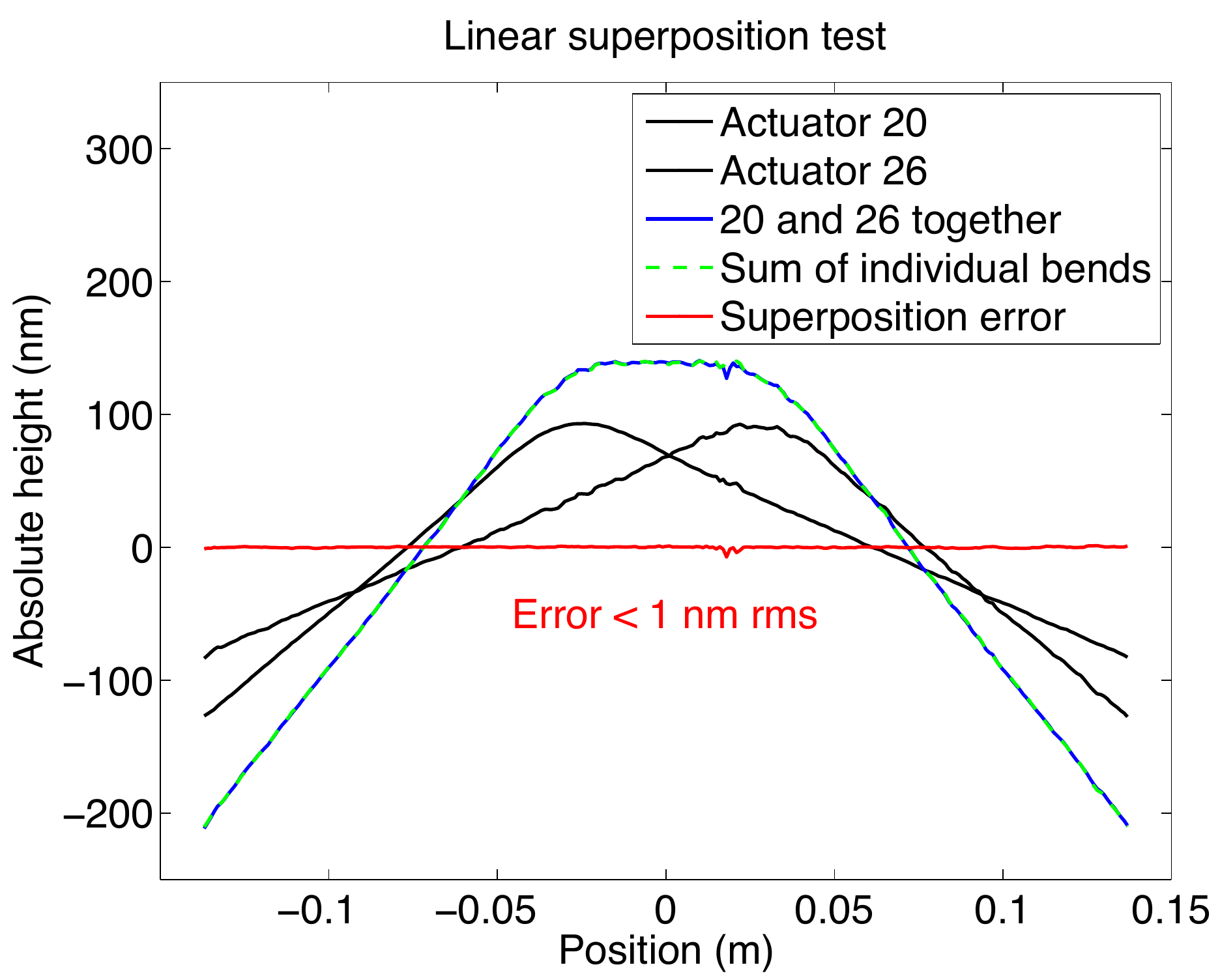}
}
\caption{The XDM obeys linear superposition, allowing use of a matrix
  equation to relate figure and voltage. In this case the result of
  bending actuators 20 and 26 at the same time is nearly identical to
  the sum of the measurements when moving them
  individually. Measurements done in ''center'' position only.}
\label{fig:fig_super}
\end{figure}

We did several experiments to verify that this was the case in the 
XDM as built. As shown in Figure~\ref{fig:fig_super}, superposition
holds very well. 
In this case we commanded first actuator 20, then actuator 26 
above bias individually. Then we commanded them at the same time.
We evaluated the change in height by subtracting from each a measurement at bias
voltage. The actual measurement of actuators 20 and 26 commanded together is 
very nearly the same as the sum of the two individual measurements.

The second aspect of linearity is that if we scale the input
$\mathbf{v}$ by a constant, then the output is scaled by the same
constant. The response of the XDM to voltage is close enough to linear
that this assumption is valid.  Since hysteresis only about 1\%, we
can treat the XDM as a linear system and use the matrix equation to
control it. The validity ignoring hysteresis and assuming perfect
linearity will be studied in future work.

\subsection{Fitting the height with a matrix or non-linear optimization}

Now that we have determined that the matrix $\mathbf{H}$  describes
our actual XDM, we can use such a matrix approach.  The 
final question is how do we actually implement the solution to the inverse problem. In
adaptive optics, the standard approach~\cite{Hardy}
is to calculate the pseudoinverse of $ \mathbf{H}$
and estimate the voltages with
\begin{equation}
\label{ }
\hat{\mathbf{v}} = \mathbf{H^+}\mathbf{\phi}.
\end{equation}
The pseudoinverse is usually calculated with a singular value decomposition (SVD).
However, care must be taken in the pseudoinverse process to reject
very small singular values (for example, see the discussion in Gavel~\cite{GavelWaffle}
on this problem and possible better approaches).

Characteristics of the XDM make the SVD approach workable, but only
with caution. In particular the broadness of the influence function
means that there is a huge dynamic range variation with spatial
frequency. While the XDM can make several microns of cylinder, it can
make only a few nanometers of the highest spatial frequencies. When
calculating the SVD, the singular values for $\mathbf{H}$ span a range
of over 100,000. If all singular values are included when the
pseudoinverse is calculated, huge noise inflation can occur. However,
if too many singular values are suppressed, very little of the height
shape will be correctly fitted. Through an analysis of different
levels, we have determined that the best tolerance results in keeping
the 21 largest singular values and those modes in the SVD. As a result
we correct only about half of the full frequency range possible on the
XDM, up to a spatial period of 4 cm.

In practice the SVD works well, but due to rejecting just over half of
the modes, we wanted to explore other options. At the present
computational costs are not a significant factor, so we explored
optimization methods. These have been used elsewhere in
AO~\cite{Pueyo2009Optimal-dark-ho} to control DMs. We implemented a
variety of optimization methods with functions in Matlab's
Optimization Toolbox. Simulations with model influence functions were
used to study several options, including Matlab's linear programming
method $\tt{linprog}$ to minimize either the L1 norm, the L-infinity
norm or to minimize the actuator stroke. We also used Matlab's
quadratic programming method $\tt{quadprog}$ to minimize the L2
norm. Of these options, the quadratic programming method worked the
best. In simulations it produced the least figure error and did not
have issues with convergence. We implement the L2-norm optimization
with an initial estimate of the voltages obtained with use of the
psuedoinverse matrix, and then use the option $\tt{active}$-$\tt{set}$
and constraint the actuator voltages to change by no more than 10\% of
the total range.  Both of these methods work well and give us a way to
convert of precision metrology to actuator commands.

\section{Flattening of the deformable mirror}

The tools described above allow us to calculate voltages that 
reduce the controllable figure error on the XDM surface. Our requirement 
is to flatten the XDM to the same level of controllable error as before
assembly (3.5 nm). Our goal is to flatten it to better than 1 nm rms.

One approach to flattening the XDM is to take a single stitched measurement of 
its full figure, determine new voltages from that measurement, and apply them.
In practice, this approach does not achieve our goal.
This is due to either errors in the model, or non-linear effects such
as hysteresis. Such an ``open loop'' approach will 
be explored in future work.
The second approach is to try a ``closed loop'' control where we 
take a series of measurements, each time feeding back the residual error
and integrating it. This approach overcomes hysteresis and some non-linear effects,
and we have found it to be reliable and stable.

\subsection{Flattening one position}

For correcting a sub-section of the XDM (e.g. in center position only), 
this whole operation proceeds rapidly. The XDM is initially placed at bias voltage.
Given a calibrated measurement,
piston and tilt are removed. This residual is sent to an integral controller 
with gain 0.5 and memory 0.999. Because the interferometer view is smaller than
the XDM's length, we extrapolate the signal beyond 
the viewing area to the full 45 cm, in the process minimizing XDM curvature. 
This modified height vector $\mathbf{\phi}$
is then used for the inverse problem to estimate voltages, which are applied to 
the XDM. A new calibrated measurement is taken and the process repeats.
This converges to better than 1 nm figure error in five steps or fewer.
We can typically achieve between 0.5 and 0.6 nm rms controllable figure
over a 20 cm length section of the XDM, with an occasional best 
correction down to 0.45 nm or below.

\subsection{Flattening the entire length}
\label{sec:fullflat}

Flattening the entire length is  more of an experimental challenge. 
Changing position requires physically moving the XDM about 8 cm, 
adjusting the fringes on the interferometer, and waiting for the 
air in the enclosure to settle. More problematic is that every time the 
XDM is moved, there is the potential for changing its figure. The
connection cables off the back are numerous and quite thick, and moving
them can change the surface figure (as is easily seen in a real-time change of
the fringes on the interferometer display). The cables
 are draped on a smooth stand to reduce forces on them when the mount is translated.

Once the repeatability challenges with moving the XDM have been
overcome, we still have a question of time efficiency. We could do the
same closed-loop approach as above for one position, except taking
three measurements each time. This method would be very labor and time
intensive. Instead we flatten the XDM as above for the center
position, and then move to the right position.  In right position, the
portion of the XDM that was corrected in center is still extremely
flat, while the portion near the mirror's edge that was out of the
center field of view is uncorrected. To correct only this new portion,
and ensure that we do not introduce error where the views overlap, we
align the residual measurement so that the previously-corrected
portion has no tilt. To combine this with the previous shape on the
XDM, we take the previous voltages and estimate the correction through
multiplication with the $\mathbf{H}$ matrix. We then add the residual
to this; since the previously corrected portion has no tilt, this adds
nothing to the unmeasured portion of the XDM. Only the new portion
seen in right position has a non-zero residual. We then solve the
inverse problem (as described above) to obtain the new voltages.

This process in essence changes the voltages so that the previously uncorrected part
of the right view is corrected, while preserving the flat surface shape in the center section.
In practice we usually use the quadratic programming optimization and 
the result is under 1 nm rms figure error in less than five iterations. 
Once the right position is flattened, we move to
left and correct that remaining portion in a similar manner.

At the end of this process we have a complete voltage set, which we place on the
XDM and measure at each of the three positions. The figure error is typically
under 2 nm. We then use this stitched calibrated measurement to update
the voltages. We estimate the correction shape made by the entire mirror by taking
the voltages and multiplying by the  $\mathbf{H}$ matrix. We then add the
stitched residual, and solve the inverse problem to fit this new height. Typically one or just
two iteration of this is all that is necessary to produce a controllable figure error of less
than 1 nm rms.

\begin{figure}[htbp]
\centerline{
\includegraphics[width=\columnwidth]{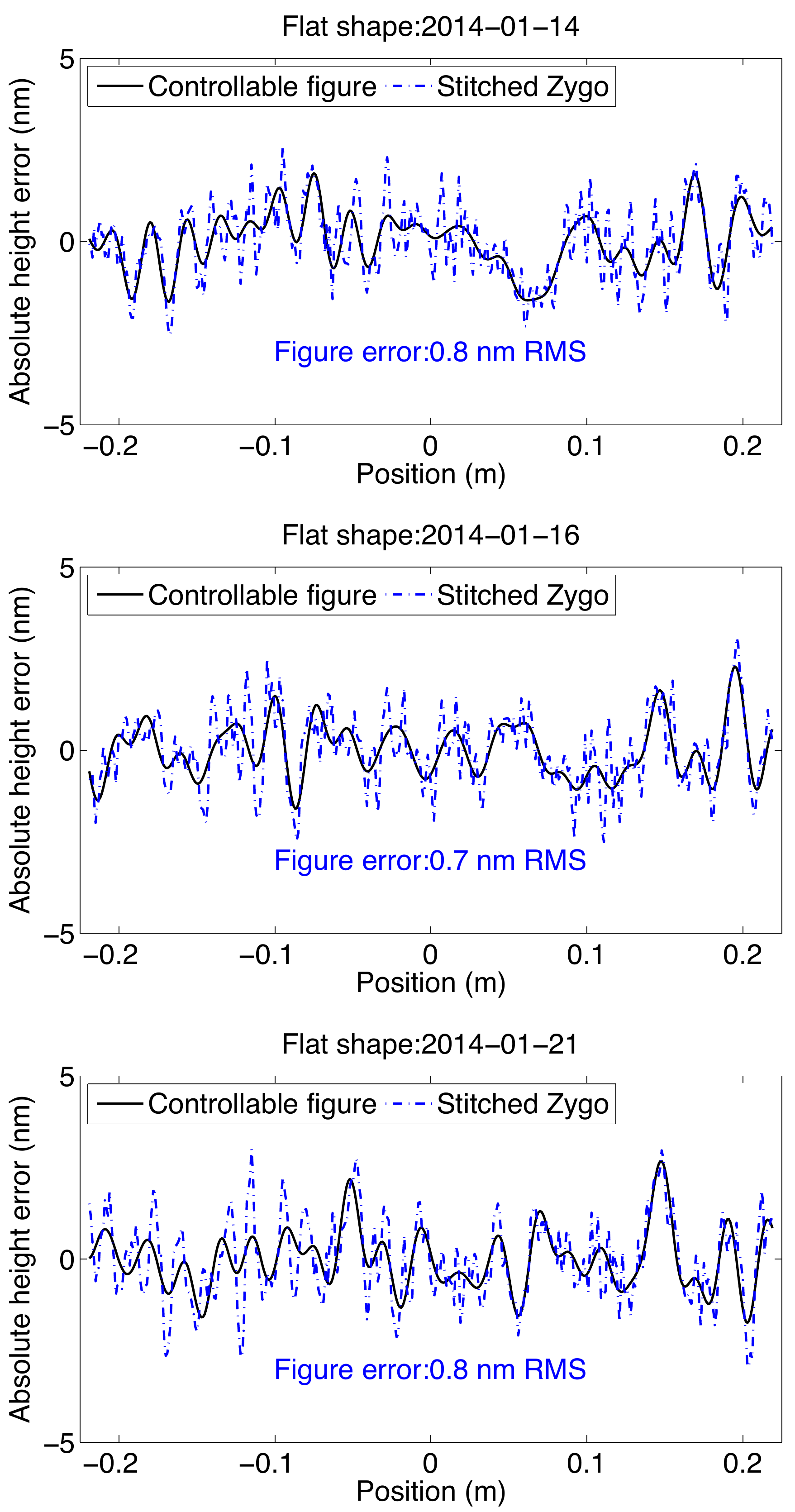}
}
\caption{In three separate experiments we achieved better
than 1.0 nm rms controllable figure error. Three plots show the stitched calibrated measurements
for the full XDM length, along with the controllable portion.
}
\label{fig:fig_flat}
\end{figure}

To demonstrate this process we have executed it three times, producing 
controllable figure errors of under 1 nm rms. The three results are shown
in Figure~\ref{fig:fig_flat}. 
All runs occurred in January 2014. 
On the 14th, we achieved 0.8 nm rms controllable figure error. 
On the 16th, we achieved 0.7 nm rms controllable figure error. 
On the 21st, we achieved 0.8 nm rms controllable figure error. 
These are calculated across a 43.8 cm length on the XDM. As discussed 
below, these three trials represent different realizations of the same
fundamental flattening process in the presence of noise.

Sub-sections of these measurements are even flatter. In the January 16th measurement
 a 21.6 cm section from x positions -8 cm to 
13.6 cm has 0.5 nm rms controllable figure error. This number is also
readily achievable in flattening a 20-cm section of a single view of the XDM, 
as described in the 
previous section.

\begin{figure}[htbp]
\centerline{
\includegraphics[width=\columnwidth]{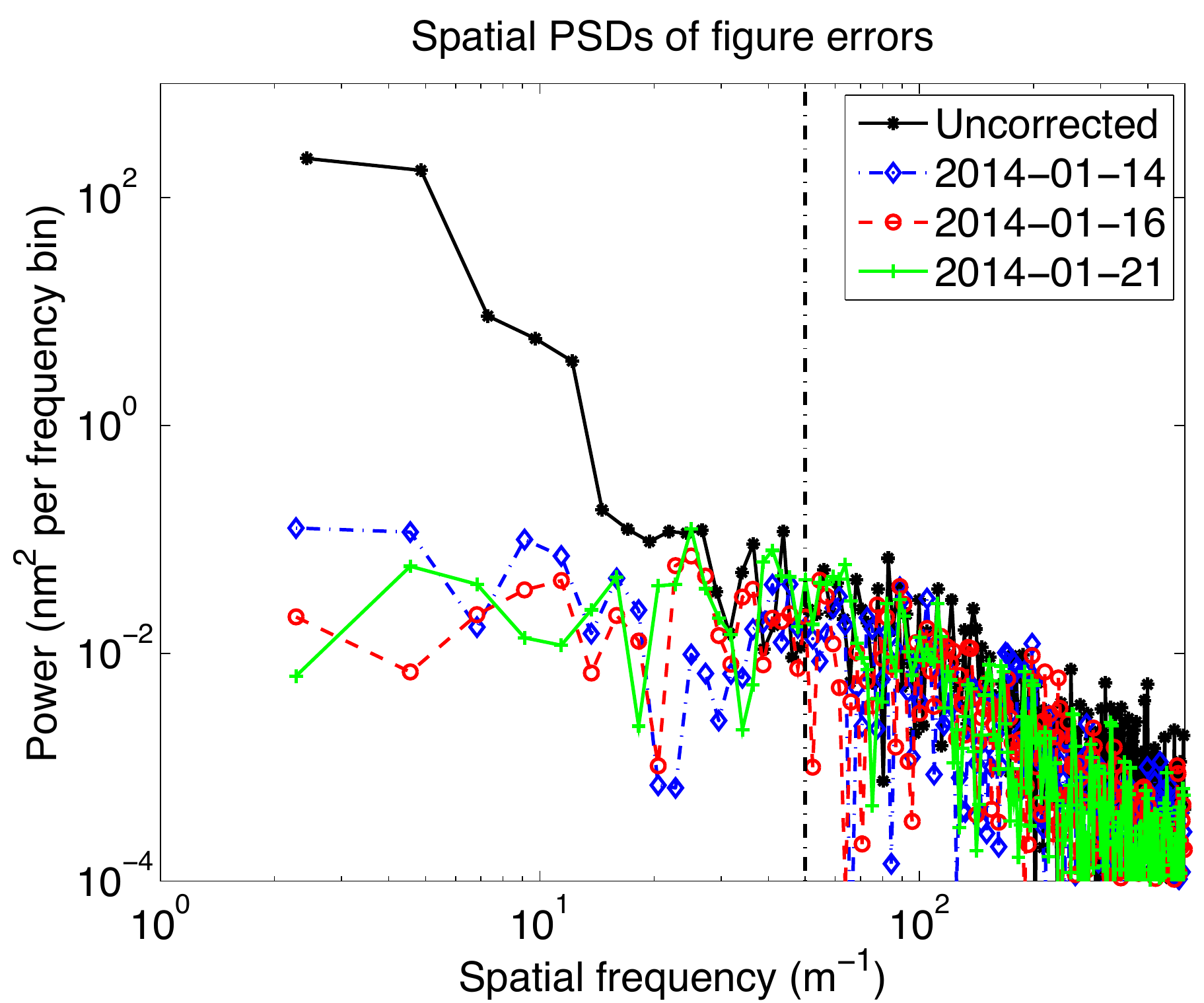}
}
\caption{Spatial power spectral densities of the 
figure at bias and the flattening residuals show that all three trials
have a similar distribution of residual error, and that we are correcting
figure to about half of the XDM's maximum controllable frequency  (shown
by the dashed vertical line).}
\label{fig:fig_flatpsd}
\end{figure}

The three residual figure errors shown in Figure~\ref{fig:fig_flat} all look different,
indicating that we have no static error source that is limiting correction.
We can further analyse the results by estimating the spatial power spectral densities (PSDs)
through the modified periodogram method~\cite{DSP} 
(i.e. $\tt{periodogram}$ with a Hanning window in Matlab).
As shown in Figure~\ref{fig:fig_flatpsd}, the three trials all have
similar power through the controllable region. 
We are not significantly
correcting modes with periods shorter than 4 cm, which is consistent
with the limitations of the inversion methods and the single-position flattening results
mentioned above.

\begin{figure}[htbp]
\centerline{
\includegraphics[width=\columnwidth]{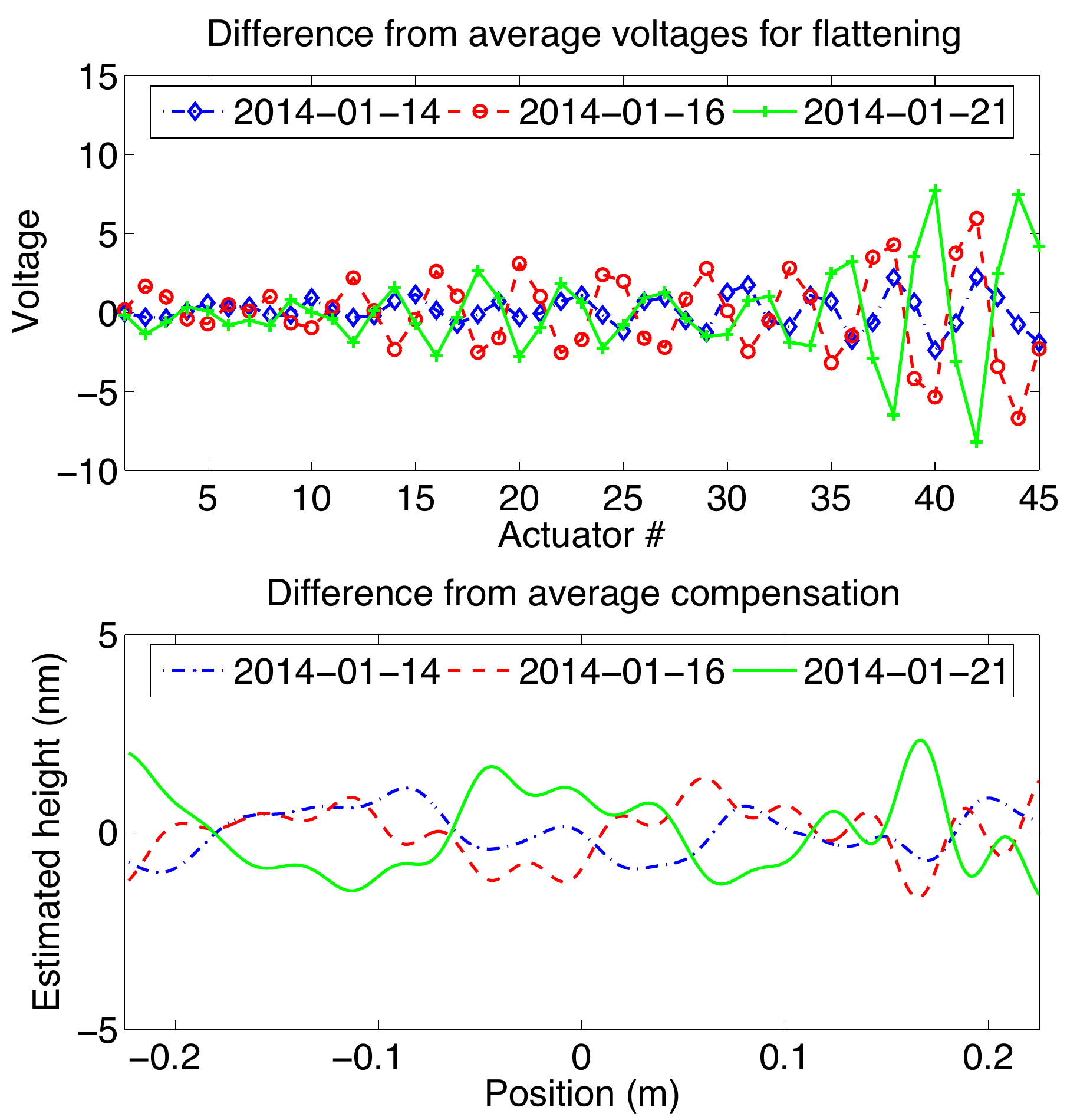}
}

\caption{The final voltages (top) obtained in the three experiments
  differ in some ways; up to about 8 V from the average of the three
  voltage sets. However, the estimated figure compensation (bottom)
  produced by these commands is very similar. The difference from the
  average of the three is less than 1 nm rms for all three trials,
  indicating stability of the overall XDM figure error and robustness
  of the flattening process and measurements.}
\label{fig:fig_voltages}
\end{figure}

Each of the three experiments was conducted from an initial condition
of bias voltage for all actuators. The difference final voltages
applied to the XDM from the average voltages for the three trials are
shown in the top of Figure~\ref{fig:fig_voltages}.
Though the low numbered actuators all track extremely well, there is
difference in the middle and especially high numbers. However, these
differing voltage sets produce nearly the same shape, as shown with
the difference from average compensation at bottom in the Figure.  The
estimated compensation was calculated with the $\mathbf{H}$ matrix.
Different voltages producing nearly the same figure is possible for
two reasons. First, the voltages represent curvature, not position.
Second, the actuators near the edge bend the XDM very little and
translate into small figure changes.  The shape made the XDM is
estimated to vary by only a few nanometers across the entire
length. This points to the stability of the overall figure error on
the XDM and robustness of our algorithms and experimental procedures.

At this time our largest error sources are intrinsic to the
experimental setup. These are the small changes in figure as the
temperature changes with time during the trials, and any small
distortions of the XDM surface as the mount is moved and forces on the
cables change. Even with these errors, we can still reliably and
repeatably achieve sub-nanometer flattening of the entire XDM length
from zero initial conditions.

These surface figure measurements are done with visible light
metrology and are not directly comparable to an at-wavelength focusing
test, such as that conducted by Mimura et
al.~\cite{Mimura2010Breaking-the-10}. Final focus spot size will
depend not only on the figure of the XDM, but on the x-ray wavelength,
F-number of the final focusing optic, the incident graze angle, as
well as the net sum rms figure error of all the optics

Rayleigh's criterion implies near diffraction limited focus is
possible when the optical path difference over the aperture is $<$
(rms figure error/16)/grazing incidence angle. At 10 keV energy, this
corresponds to an rms figure error $<$ 0.008 nm/graze angle. Assuming
the mirror used at 1 milli-radian graze angle (typical for mirrors
this long), this corresponds to surface normal figure error of $<$ 8
nm rms. Hence, the sensitivity of our surface normal mirror correction
is more than adequate to meet this requirement.

In summary, we have flattened our XDM to a flatter figure than the
best previously published results (detailed above in
Section~\ref{sec:intro}) for a deformable mirror of similar length (35
cm). Our sub-nm flattening level is comparable to the best achieved by
other deformable optics, but on a substrate more than three times as
long (45 cm vs 12 cm).

\section{Conclusions and Future Work}

We have manufactured a 45-cm long x-ray deformable mirror. 
The initial substrate was polished to 3.5 nm rms figure error; 
after actuator bonding and mirror assembly this error became 19 nm rms.
We have used very precise visible-light interferometry and detailed 
characterization of the XDM to perform closed-loop control to flatten its surface.
Starting from zero initial conditions, we can reliably and repeatedly 
flatten the controllable figure of the XDM to sub-nanometer levels.
Our best correction of the full length is 0.7 nm RMS; for smaller 20-cm sections
was have achieved 0.5 nm rms. 

The next challenge is to 
maintain such a flat shape through time without the use of external metrology. 
Our XDM has 45 strain gauges (one per actuator)
and eight temperature sensors. These will be used to correct for both
temperature dependent changes in figure as well as other non-linear effects,
such as time-dependent changes in the PMN response. 
In our future work we will conduct a full characterization of the gauges and sensors.
Once calibrated, we will use them for feedback control to maintain
both the cylinder and figure of the XDM over periods of several hours. 
A secondary task is to better understand the single-step shaping of the XDM,
and whether we are limited by knowledge of the influence functions, hysteresis,
or some other factor.
Once we can reliably flatten and maintain the XDM as flat, and make
arbitrary shapes, we will move on to at-wavelength testing of the XDM
at the Advanced Light Source at Lawrence Berkeley National Laboratory.
Of particular interest are studying different wavefront sensing methods
to provide accurate and rapid {\it in situ} metrology of the XDM.

\section*{Acknowledgments}    
This work performed under the auspices of the U.S. Department of
Energy by Lawrence Livermore National Laboratory under Contract
DE-AC52-07NA27344.  The document number is LLNL-JRNL-649073.  Early
technology development was supported at NG-AOX through internal
research and development funding. The authors thank Brian Bauman
(optics), Carol Meyers (optimization methods), and Peter Thelin
(laboratory facilities) for their advice or assistance.  The authors
also thank the reviewers for their diligent evaluation and helpful
comments that have improved this paper.



\end{document}